\documentstyle[fleqn]{article}

\newcommand{\be}{\begin{equation}}
\newcommand{\ee}{\end{equation}}
\newcommand{\bea}{\begin{eqnarray}}
\newcommand{\eea}{\end{eqnarray}}
\newcommand{\bean}{\begin{eqnarray*}}
\newcommand{\eean}{\end{eqnarray*}}
\newcommand{\sslash}[1]{\not{\!#1}}
\newcommand{\lslash}[1]{\not{\!\!#1}}

\begin{document}

\noindent
{\Large Twist - 3 polarized structure functions~
           \footnote{Talk presented by J. Kodaira at the QCD 99 Euroconference,
        Montpellier, July 7-13, 1999.}}

\vspace{2ex}
\noindent
Jiro Kodaira$^{a}\!\!$~
        \footnote{Research supported in part by the Monbusho Grant-in-Aid
                  for Scientific Research No.C(2)09640364.}
and Kazuhiro Tanaka$^b$

\vspace{2ex}
\noindent
$^a$Dept. of Physics, Hiroshima University, 
     Higashi-Hiroshima 739-8526, Japan

\vspace{2ex}
\noindent
$^b$Dept. of Physics, Juntendo University,
                          Inba-gun, Chiba 270-1606, Japan 

\vspace{4ex}
We review the nucleon's twist-3 polarized structure functions 
from the viewpoint of gauge invariant, nonlocal light-cone operators in QCD.
We discuss a systematic treatment of
the polarized structure functions and the corresponding
parton distribution functions. 
We emphasize unique features of higher twist distributions,
and the role of the QCD equations of motion to derive
their anomalous dimensions for $Q^{2}$-evolution. 
\vspace{4ex}

\noindent
{\bf 1. Introduction} 

\vspace{12pt}

In the last ten years, great progress has been made
both theoretically and experimentally in hadron spin physics.
Furthermore, in conjunction with 
new projects like the \lq\lq RHIC spin project\rq\rq, \lq\lq polarized
HERA\rq\rq , etc.,  we are now
in a position to obtain more information on the 
spin structure of nucleons.

In the case of spin-dependent processes, 
it has been known that 
the twist-3 contributions can be measured as leading effects
in certain asymmetries.
Since the higher twist effects describe
the coherent quark-gluon behavior inside the nucleon,
we could obtain information on the correlation of
quarks and gluons beyond the (QCD) parton model (leading twist effect).

In this talk, we will concentrate on the twist-3 polarized structure
functions~\cite{beli}.
We present a systematic treatment of these functions
in terms of gauge invariant, nonlocal light-cone operators, 
and summarize the recent theoretical progress on
the QCD evolutions~\cite{kodatana}.

\vspace{12pt}
\noindent
{\bf 2. Classification of twist-3 structure functions}
\vspace{12pt}

To describe a variety of high-energy processes in a universal language,
it is desirable to have a definition of parton distribution
functions based on the operators in QCD.
The traditional approach relies on 
the operator product expansion (OPE), 
but it can be applied only to a limited
class of processes.
This calls for an approach based on the {\it factorization}
as a generalization of the OPE.

\vspace{12pt}
\noindent
{\bf 2.1. Parton distributions}

It is instructive, at first, to consider the deep inelastic scattering
(DIS) in free field theories.
The result provides a guide to
operator definitions of the structure functions
(parton distributions).
As is well known, the relevant quantity for the DIS
is the hadronic tensor:
\be
  W_{\mu\nu}
 = \int \frac{d^{\,4} z}{2\pi} e^{- i q\cdot z} \langle P S |
               [J_{\mu}(0)\,,\,J_{\nu}(z)]| P S \rangle \ .\label{ht}
\ee
Here $|PS\rangle$ is the hadron state with momentum $P$ and spin $S$,
$q^{\mu}\,(Q^2 = - q^2 )$ is the momentum transferred to the hadron, and 
$J_{\mu}$ is the hadron's electromagnetic (weak) current.
In the Bjorken limit, the hadronic tensor (\ref{ht}) is governed by
the behavior of the current product near the light-cone
(light-cone dominance).
In the free field theory, the current product can be calculated to be
\be
 [ J_{\mu} (0) \,,\, J_{\nu} (z) ] =
  - \left(\partial^{\alpha} \Delta (z )\right)
   [ S_{\mu \alpha \nu \sigma} 
           {\cal U}_V^{\sigma} ( 0 , z) - i 
        \epsilon_{\mu\alpha\nu\sigma}
           {\cal U}_A^{\sigma} ( 0 , z) ] , \label{ccinfree}
\ee
up to irrelevant terms to the DIS.
$\Delta(z)$ is the commutator function,
$S_{\mu \alpha \nu \sigma}\equiv  g_{\mu\alpha} g_{\nu\sigma}
- g_{\mu\nu} g_{\alpha\sigma} + g_{\mu\sigma} g_{\nu\alpha}$,
and the {\it nonlocal operators} are defined by
\bea
  {\cal U}_V^{\sigma} (0 , z) &\equiv& 
     \bar{\psi} (0 ) \gamma^{\sigma} \psi (z)
              - \bar{\psi} (z) \gamma^{\sigma} \psi (0 ), \label{bico0}\\
  {\cal U}_A^{\sigma} (0 , z) &\equiv& 
     \bar{\psi} (0) \gamma^{\sigma} \gamma^5 \psi (z)
   + \bar{\psi} (z) \gamma^{\sigma} \gamma^5 \psi (0) . \nonumber
\eea
The singularity of $\Delta(z)$ (or more generally the light cone
dominance) selects
the integration region with $\eta \sim z_{\perp} \sim 0$
when we write $z^{\mu}$ as
$z^{\mu} = \eta p^{\mu} + \lambda w^{\mu} + z_{\perp}^{\mu}$ using
two auxiliary light-like vectors $p^2 = w^2 = 0 \, , \,p \cdot w = 1$.
(We take 
$P^{\mu} = p^{\mu} + (M^{2}/2) w^{\mu}$
and $q^{\mu} = -x p^{\mu} + \left( Q^{2}/2x + 
M^{2}x / 2 \right) w^{\mu} + {\cal O}( 1 / Q^{2})$
with the hadron mass $M$ and the Bjorken variable
$x \equiv Q^{2} / (2P\cdot q)$.)
Therefore, we can expand the matrix element of 
(\ref{ccinfree}) in powers of the deviation 
from the light-cone, $z^{\mu} -\lambda w^{\mu}$, and approximate it as
\[ \langle P S | {\cal U}_{V,A}^{\sigma} (0 ,z) | P S \rangle \simeq
    \langle P S | {\cal U}_{V,A}^{\sigma} (0 , \lambda w ) 
               | P S \rangle .\]
By parameterizing the first term of (\ref{bico0}) as
\be
 \langle PS | \bar{\psi} (0 )
       \gamma^{\sigma} \psi (\lambda w )| PS \rangle
    = 2 p^{\sigma} \hat{q} (\lambda )
  + 2 w^{\sigma} M^2 \hat{f}_4 (\lambda )    , \label{eq:decomp}
\ee
and substituting it into (\ref{ht}), 
the unpolarized structure function (quark distribution) is
defined as the Fourier transform of $\hat{q}(\lambda)$ 
into the Bjorken-$x$ space,
\be
 q(x) = \int \frac{d\lambda}{4\pi} e^{i\lambda x}
  \langle P S| \bar{\psi} (0)\sslash{w}\, \psi (\lambda w)| P S\rangle 
  . \label{nsq}
\ee
The second term of (\ref{eq:decomp}) 
is the twist-4 contribution.
Other terms of (\ref{ccinfree}) can be treated similarly.
These results yield the 
{\it factorization} into the $c$-number coefficients (short-distance part) 
and the parton distributions (long-distance part). 

In the presence of QCD interaction,
the above results are modified in two respects.
First, the higher order interactions
produce the logarithmic ($\ln(Q^{2}/\mu^{2})$)
corrections to $\Delta (z)$, and correspondingly
the parton distribution functions acquire a dependence on the
renormalization scale $\mu$.
Second, the coupling of the ``longitudinal'' gluons replaces
$\Delta(z)$ as~\cite{BB}
\[   \Delta (z) \rightarrow \Delta(z)[0, z]  ,\]
where
\[    [y,z]= {\rm P}\! \exp \left(ig \!\! \int_0^1 \!\!\! dt
      (y-z)_\mu A^\mu(ty+(1-t)z) \right)\  \]
is the path-ordered link operator
connecting the points $z_{\mu}$ and $y_{\mu}$ along the straight line.
By absorbing this factor,
the nonlocal light-cone operators which define the parton
distribution functions now preserve gauge invariance.

\vspace{12pt}
\noindent
{\bf 2.2. Twist-3 quark distributions}

Now, it is not difficult to define the parton
distribution functions in QCD~\cite{CS}.
We consider the following quantity involving the nucleon matrix element:
\be
 \int_{-\infty}^{+\infty} \frac{d\lambda}{2\pi} e^{i\lambda x} 
        \langle P S|
       \bar{\psi}(0) [0,\lambda w] \Gamma \psi(\lambda w) |P S \rangle 
    \ . \label{eq:gam}
\ee
Here $\Gamma$ is a generic Dirac matrix, and
the link operator $[0, \lambda w]$
makes the operators gauge invariant.
Equation (\ref{eq:gam}) defines the distribution function
for a quark with momentum $k\cdot w=x P \cdot w= x$.
$\Gamma$ can be any Dirac matrix, depending on which hard
process is considered.
An important observation made in~\cite{JJ}
is that one can generate all quark distribution functions 
up to twist-4 by substituting all the possible $\Gamma$.

By decomposing (\ref{eq:gam}) into independent tensor 
structures, one finds nine independent quark distribution functions
associating with each tensor structure~\cite{kodatana}.
We list here only the twist-3 polarized distribution functions.
\bean
  \int { d\lambda  \over 2 \pi } e^{i\lambda x} \langle PS |
  \bar{\psi}(0) [0, \lambda w]\gamma^{\mu}\gamma_5 
  \psi(\lambda w) |PS \rangle
 &=& 2 \ g_{T}(x, \mu^2) S_{\perp}^{\mu} + \cdots \ , \\
  \int { d\lambda  \over 2 \pi } e^{i\lambda x} \langle PS |
  \bar{\psi}(0) [0, \lambda w]\sigma^{\mu\nu}i\gamma_5 
  \psi(\lambda w) |PS \rangle
  &=& 2 \ h_{L}(x, \mu^2)M(p^\mu w^\nu - p^\nu w^\mu )(S \cdot w)
   + \cdots \, ,
\eean
where $S^{\mu} =  S_{\parallel}^{\mu} + S_{\perp}^{\mu}$
with $S_{\parallel} \cdot w = S\cdot w$.

Three comments are in order~\cite{kodatana}.
(1) Since the definition of the twist here
is based on the simple power counting with respect to $1/Q$,
there is a slight mismatch with the conventional definition
as ``dimension minus spin'' of the relevant operators.
Actually both distributions $g_{T}$ and $h_{L}$ contain the
contributions from the twist-2 pieces.
(2) $g_{T}$ and $h_{L}$ have different chiralities
(``even'' and ``odd'').
(3) The light-cone quantization formalism~\cite{JJ,KS}
is a conceptually useful approach to the twist counting. 
In this approach, quark fields are decomposed into ``good''
and ``bad'' components.
The good component represents an independent
degree of freedom;
the bad components are not dynamically independent
and can be reexpressed by a coherent quark-gluon pair.
Therefore, only the twist-2 distributions are literally
the ``distributions'' and correspond to the parton model,
and the higher twist distributions are the multiparton 
(quark-gluon) correlations.

\vspace{12pt}
\noindent
{\bf 2.3. Twist-3 gluon distributions}

Our analysis can be extended to the gluon distribution functions.
The gauge-invariant definition of the gluon distribution functions
is provided by~\cite{CS,mano}
\[ \int \frac{d\lambda}{2\pi} e^{i\lambda x}
   \langle PS| w_{\alpha} G^{\alpha \mu}(0) [0, \lambda w]
  w_{\beta} G^{\beta \nu}(\lambda w) |PS \rangle
  = - \frac{x}{2} \, {\cal G}_{3T}(x, \mu^{2})  
   i \epsilon^{\mu \nu \alpha \beta} S_{\perp \alpha}w_{\beta}
    + \cdots \ ,\]
where $G^{\mu \nu}$ is the field strength tensor,
and we show the twist-3 term explicitly.
The gluon distributions mix through renormalization
with the flavor singlet chiral-even quark distributions.
On the other hand, there exists no gluon distributions
that mix with the chiral-odd quark distributions.

\vspace{12pt}
\noindent
{\bf 2.4. Twist-3 three-particle distributions}

Coherent many-particle contents of the nucleon are described by
multiparton distribution functions.
The twist-3 quark-gluon
correlation functions are defined as
\be
 \int \frac{d\lambda}{2\pi}\frac{d\zeta}{2\pi}
   \,e^{i\lambda x + i\zeta (x'-x)}
  \langle PS|\bar{\psi}(0) \Gamma [0,\zeta w]
  \, g\, G^{\mu\nu}(\zeta w)
     [\zeta w,\lambda w] \psi(\lambda w)|PS\rangle .\label{threeb}
\ee
Similarly to the quark distributions, one can define the  
multiparton distributions by considering possible Dirac matrices for
$\Gamma$.
The treatment here can be extended to the case of 
three-gluon correlation functions, which are relevant to 
the (singlet) quark distribution $g_{T}(x)$ and the gluon distribution
${\cal G}_{3T}(x)$.
Although the extension is straightforward,
we do not go into the details of the three-gluon
nonlocal light-cone operators~\cite{BKL,kntty}.

\vspace{12pt}
\noindent
{\bf 3. QCD evolutions}
\vspace{12pt}

The calculation of $Q^2$ evolutions for the higher twist terms
is generally very complicated due to the presence of multiparton
distributions.
Although it is possible to generalize the DGLAP approach
to the three-body case and calculate its evolution kernel,
one is forced to use some particular techniques, e.g.
the use of the light-like axial gauge etc.~\cite{BKL,DMu}
In this talk, we report an approach based on the ``local composite 
operators'' in a ``covariant'' gauge~\cite{KODS}.
The advantage of this approach is that all the relevant steps
can be worked out based on the standard and familiar field theory
techniques.
Therefore, the calculation is very straightforward and
we can assess unambiguously the necessary information for the
$Q^2$ evolutions.

\vspace{12pt}
\noindent
{\bf 3.1. Local operators} 

By taking the moment of distribution functions,
one can show that there is one-to-one correspondence between  
the nonlocal and the local operators.
The simplest example is provided by
the unpolarized twist-2 quark distribution
(\ref{nsq}):
\bean
  \int dx x^{n-1} q (x) &=& \frac{1}{2} \int d \lambda 
          \left\{ \left( \frac{\partial }{ i \partial\lambda } \right)^{n-1}
             \delta (\lambda ) \right\} \langle P S| \bar{\psi} (0) \sslash{w}
     [0, \lambda w] \psi (\lambda w)| P S \rangle \\
    &=&  \frac{1}{2} \ \langle PS | \bar{\psi} (0) \sslash{w}
     \left( i w\cdot D \right)^{n-1} \psi (0) | P S\rangle \ .
\eean
This equation implies that
the moments of the parton distributions 
are given by the matrix elements of the corresponding
gauge-invariant, local composite operators.
For the twist-3 distributions,
the situation is the same except that one should
take into account 
the ``double'' moments for the three-body distributions.

\vspace{12pt}
\noindent
{\bf 3.2. Operator relations} 

The common feature for the higher twist operators is that
there appear a set of local operators
with the same quantum numbers, the number of which increases with
spin (moment), 
and these operators are not all independent
but related through the QCD equation of motion~\cite{P80,SV}.

One of the most convenient method to identify the relevant operators
and a relation among them is to utilize the exact operator
identities satisfied by the gauge invariant nonlocal operators.
In the case of non-singlet part of $g_{T}$, 
the following operator identity can be
obtained by explicit differentiation~\cite{BB}.
\bean
  \lefteqn{ z_{\mu} \left(
     \frac{\partial}{\partial z_{\mu}}
        \bar{\psi}(0) \gamma^{\sigma} \gamma_5 [0, z] \psi( z )
      -  ( \sigma \leftrightarrow \mu ) \right)}\\
    &=& \!\! \int_0^1 dt \bar{\psi}(0) [0, tz] \sslash{z} \left\{
          i \gamma_5 \left( t - \frac{1}{2} \right) g G^{\sigma\rho}(tz)
           z_{\rho} -  \frac{1}{2} g \tilde{G}^{\sigma\rho}(tz) z_{\rho} \right\} 
     [tz , z] \psi ( z ) \\
   & & + \, 2 \, m_{q} \bar{\psi}(0) \gamma_5 \sigma^{\sigma\rho} z_{\rho}
       [ 0 , z] \psi (z )\\
   & & + \, \bar{\psi}(0) \gamma_5 \sigma^{\sigma\rho} z_{\rho} [ 0 , z] E
         \psi ( z ) - \psi (0) \overleftarrow{E} 
          \gamma_5 \sigma^{\sigma\rho} z_{\rho} [ 0 ,z] \psi ( z ) .
\eean
Here $E = i \lslash{D} - m_{q}$ is the equation of motion operator
and $m_{q}$ represents
the quark mass generically.
This identity is exact through twist-3.
Taylor expanding the above equation around $z_{\mu} = 0$,
we easily identify the local operators, which contribute to the moment
of $g_{T}$, and also
the relation among them.
\bean
  R_{n,F}^{\sigma} &=&
     - i^{n-1} {\cal SAS} \overline{\psi}
       \gamma_5 \gamma^{[ \sigma} D^{\{\mu_1 ]} \cdots D^{\mu_{n-1}\}}
          \psi ,\\
  R_{n,l}^{\sigma} &=& 
     \frac{1}{2n}\left(V_{l}-V_{n-1-l} + U_{l} + U_{n-1-l} \right) ,\\
  R_{n,m}^{\sigma} &=& - i^{n} {\cal S}\,m_{q}
          \overline{\psi}\gamma^{\sigma}\gamma_5
           D^{\mu_{1}}\cdots D^{\mu_{n-2}}\gamma_{\mu_{n-1}}
          \psi \ , \\
  R_{n,E}^{\sigma} &=& i^{n-2}\frac{n-1}{2n}{\cal S} \bar{\psi} \left[  E
   \gamma^{\sigma} \gamma_5 D^{\mu_{1}} \cdots D^{\mu_{n-2}}
      \gamma^{\mu_{n-1}} +  \gamma^{\sigma} \gamma_5
    D^{\mu_{1}} \cdots D^{\mu_{n-2}}\gamma^{\mu_{n-1}} E  \right] \psi \ ,
\eean
where ${\cal S}$ symmetrizes over $\mu_{i}$ and ${\cal A}$
antisymmetrizes $\sigma$ and $\mu_{1}$, and 
\bean
   V_{l}&=& - i^{n} g {\cal S} \bar{\psi} 
       D^{\mu_1} \cdots G^{\sigma \mu_l} \cdots
       D^{\mu_{n-2}}\gamma^{\mu_{n-1}}\gamma_{5} \psi \ ,\\
   U_{l}&=& - i^{n-1} g {\cal S} \bar{\psi}
      D^{\mu_1} \cdots \tilde{G}^{\sigma \mu_l} \cdots
      D^{\mu_{n-2}}\gamma^{\mu_{n-1}} \psi \ .
\eean
The $R_{n,l}$ is related to the double moment
of quark-gluon three-body correlation functions (\ref{threeb})
with $\Gamma = \gamma_{\mu}, \gamma_{\mu}\gamma_{5}$. 
These operators satisfy,
\[   R_{n,F}^{\sigma} = 
         \frac{n-1}{n} \,R_{n,m}^{\sigma}
             + \sum_{l=1}^{n-2} (n-1-l)
                 R_{n,l}^{\sigma} +  R_{n,E}^{\sigma} \ .\]

For the singlet part, the situation becomes much more complicated.
We have additional operators that are made of two or three gluon fields, 
as well as the BRST invariant, gauge variant operators.
The explicit expressions for the operators and the derivation
of the identity satisfied by them are found in Ref.~\cite{kntty}.

For other distributions of twist-3 and higher,
we can follow exactly the same procedure.
In the case of e.g. $h_L$, from the quantity,
\[   \frac{\partial}{\partial z_{\mu}}
  \left\{ \bar{\psi}(0) i \gamma_{5} \sigma_{\mu \nu} z^{\nu} [0, z]
  \psi (z) \right\} \ ,\]
one can generate towers of identities for the corresponding local
operators~\cite{kodatana,JJ,KT}.
 
\vspace{12pt}
\noindent
{\bf 3.3. Calculations}

Since we identify a complete set of operators,
the next step is to calculate the anomalous dimensions 
(renormalization constants) for these operators to derive
the $Q^2$ evolution.
For this task, the following theorem~\cite{JL76} is very helpful.
This theorem tells us that in non-abelian gauge theory,
three kinds of operators will mix under 
the renormalization of gauge invariant operators.
(I)the gauge invariant operator itself.
(II)the BRST invariant operators.
(III)the operators which are proportional to the equation of 
motion (EOM).
These EOM operators involve both BRST ``invariant'' and
``variant'' ones~\cite{KODS,KOD5}.
Although the physical matrix elements of the BRST
invariant and EOM operators
vanish~\cite{kodatana}, it is necessary to consider these operators to
complete the renormalization, in other words to obtain the correct
renormalization constants for the gauge invariant operators.

At the lowest twist level, the above complicated operator 
mixing does not come into play,
because there exists neither an EOM nor BRST invariant
operator of twist-2. 
The EOM as well as the BRST invariant operators always have smaller
spin by at least one unit than the possible highest spin operators
of the same dimension.
At the higher twist ($ \geq 3$) level, both EOM and BRST invariant
operators participate in the renormalization mixing 
with the gauge invariant operators.
This is a characteristic feature of the renormalization for
the higher twist operators.

The explicit calculations at the one-loop level
appear in Refs.~\cite{KODS,kntty,KNT98} for $g_T$ and in Ref.~\cite{KT}
for $h_L$.

\vspace{12pt}
\noindent
{\bf 4. Conclusions}
\vspace{12pt}

We have surveyed the twist-3 polarized 
structure functions from a viewpoint of gauge invariant nonlocal 
operators.
One of the key points of our approach
is that we preserved maximal (gauge (BRST) and Lorentz) symmetries
of the theory at every step of investigation.
Another point is use of the one-to-one correspondence 
between the nonlocal and the local operator approaches
to investigate the $Q^{2}$-evolution.
We have clarified the role of the QCD equations of motion  
to reveal the interrelation between the different twist-3 distributions
and to derive their $Q^{2}$-evolution.

The polarized twist-3 distribution functions 
$g_{T}$ and $h_{L}$
are measurable as leading effects and are 
of special interest.
However, the phenomenology of twist-3 distributions is not as straightforward
as that of twist-2 distributions due to the fact that 
the twist-3 distributions are essentially
three-particle correlations: 
We need to determine many parameters to make
precise predictions.
In this respect, it is pointed out that
a drastic and universal
simplification of all twist-3 distributions occurs in the 
$N_{c}\rightarrow \infty$ limit~\cite{ABH}.
It would be a good news to solve the above problem for all practical purposes. 

We hope that various kinds of new experiments
and more detailed theoretical investigations will be able to
clarify not only perturbative but also nonperturbative
aspects of QCD related to hadron spin physics.


\begin{thebibliography}{99}
\bibitem{beli}
 A. V. Belitsky, {\it in} the XXXI PNPI Winter School on
 Nuclear and Particle Physics, (St. Petersburg, 1997) p. 369,
 hep-ph/9703432; hep-ph/9907420, and references therein.

\bibitem{kodatana}
 J. Kodaira and K. Tanaka, Prog. Theor. Phys. 101 (1999) 191.

\bibitem{BB}
 I. I. Balitsky and V. M. Braun, Nucl.\ Phys.\ {\bf B311} (1988/89) 541. 

\bibitem{CS}
 J. C. Collins and D. E. Soper, Nucl.\ Phys.\ {\bf B194} (1982) 445.

\bibitem{JJ}
      R. L. Jaffe and X. Ji, Phys.\ Rev.\ Lett.\ {\bf 67} (1991) 552.
      Nucl.\ Phys.\ {\bf B375} (1992) 527.

\bibitem{KS}
      J. B. Kogut and D. E. Soper, Phys.\ Rev.\ {\bf D1} (1970) 2901.

\bibitem{mano}
     A. V. Manohar, Phys.\ Rev.\ Lett.\ 
    {\bf 65} (1990) 2511; Phys.\ Rev.\ Lett.\ 
    {\bf 66} (1991) 289.

\bibitem{BKL}
    A. P. Bukhvostov, E. A. Kuraev and L. N. Lipatov, Sov.\ J.\ Nucl.\
    Phys.\ {\bf 38} (1983) 263; {\bf 39} (1984) 121. 
    JETP Letters {\bf 37} (1984) 483.
    Sov.\ Phys.\ JETP {\bf 60} (1984) 22.

\bibitem{kntty}
    J. Kodaira, T. Nasuno, H. Tochimura,
    K. Tanaka and Y. Yasui, Prog.\ Theor.\ Phys.\ {\bf 99} (1998) 315;
    {\it in} Proc. of 2nd Topical Workshop on Deep Inelastic 
    Scattering off Polarized Targets: Theory Meets Experiment
    (DESY-Zeuthen, 1997), DESY 97-200, p. 210.

\bibitem{DMu}
    B. Geyer, D. M\"uller and D. Robaschik, Nucl. Phys.
    Proc. Suppl. {\bf 51C} (1996) 106.\\
    D. M\"uller,  Phys.\  Lett.\  {\bf B407} (1997) 314.

\bibitem{KODS}
    J. Kodaira, Y. Yasui and T. Uematsu, Phys.\ Lett.\ {\bf B344}
    (1995) 348.\\ 
     J. Kodaira, Y. Yasui, K. Tanaka and T. Uematsu, Phys.\ Lett.\ 
     {\bf B387} (1996) 855.

\bibitem{P80}
    H. D. Politzer,  Nucl.\ Phys.\ {\bf B172} (1980) 349.

\bibitem{SV}
     E. V. Shuryak and A. I. Vainshtein, Nucl.\ Phys.\ {\bf B199}
    (1982) 951;
     Nucl.\ Phys.\ {\bf B201} (1982) 141.

\bibitem{KT}
    Y. Koike and K. Tanaka, Phys.\ Rev.\ {\bf D51} (1995), 6125. 

\bibitem{JL76}
    S.D. Joglekar and B.W. Lee, Ann. of Phys. {\bf 97}
    (1976) 160.

\bibitem {KOD5}
     H. Kawamura, T. Uematsu, J. Kodaira and Y. Yasui,
     Mod.\  Phys.\  Lett.\  {\bf A12} (1997) 135.

\bibitem{KNT98}
      Y. Koike, N. Nishiyama and K. Tanaka, Phys.\ Lett.\ 
     {\bf B437} (1998) 153.

\bibitem{ABH}
    A. Ali, V.M. Braun and G. Hiller, Phys.\ Lett.\ {\bf B266} (1991)
    117. \\
    I. I. Balitsky, V. M. Braun, Y. Koike and K. Tanaka,
    Phys.\ Rev.\ Lett.\ {\bf 77} (1996) 3078. \\
    K. Sasaki, Phys.\ Rev.\ {\bf D58} (1998) 094007.\\
    X. Ji and J. Osborne, Euro.\ Phys.\ J.\ {\bf C9} (1999) 487.

\end{thebibliography}
\end{document}